\documentclass{nature_sunnychanged}
\usepackage{graphicx} 
\usepackage{dcolumn}
\usepackage{bm}
\usepackage{amssymb,amsmath}
\usepackage{epstopdf}
\usepackage{color}

\usepackage{amsfonts}
\usepackage{amsthm,multirow,longtable}
\usepackage{epsfig}
\usepackage{comment}
\usepackage{hyperref}
\usepackage{float}
\usepackage{cite}
\usepackage{makecell}

\newcommand{\nn}{\nonumber}

\newcommand{\be}{\begin{equation}}
\newcommand{\ee}{\end{equation}}
\newcommand{\bea}{\begin{eqnarray}}
\newcommand{\eea}{\end{eqnarray}}
\newcommand{\beq}{\begin{equation}}
\newcommand{\eeq}{\end{equation}}
\newcommand{\beqn}{\begin{eqnarray}}
\newcommand{\eeqn}{\end{eqnarray}}

\newcommand{\hide}[1]{}
\renewcommand\Re{\operatorname{Re}}
\renewcommand\Im{\operatorname{Im}}

\newcommand{\mean}[1]{\langle{#1}\rangle{}}

\title{General solution to inhomogeneous dephasing and smooth pulse dynamical decoupling}

\author{Junkai Zeng$^1$, Xiu-Hao Deng$^1$, Antonio Russo$^1$ \& Edwin Barnes$^1$}
\begin{document}
\maketitle
\begin{affiliations}
\item Department of Physics, Virginia Tech, Blacksburg, Virginia 24061, USA
\end{affiliations}

\begin{abstract}
\textbf{In order to achieve the high-fidelity quantum control needed for a broad range of quantum information technologies, reducing the effects of noise and system inhomogeneities is an essential task. It is well known that a system can be decoupled from noise or made insensitive to inhomogeneous dephasing dynamically by using carefully designed pulse sequences based on square or delta-function waveforms such as Hahn spin echo or CPMG. However, such ideal pulses are often challenging to implement experimentally with high fidelity. Here, we uncover a new geometrical framework for visualizing all possible driving fields, which enables one to generate an unlimited number of smooth, experimentally feasible pulses that perform dynamical decoupling or dynamically corrected gates to arbitrarily high order. We demonstrate that this scheme can significantly enhance the fidelity of single-qubit operations in the presence of noise and when realistic limitations on pulse rise times and amplitudes are taken into account.}
\end{abstract}

In recent years, the prospect of enhanced technologies that exploit the principles of quantum mechanics has attracted great interest from many fields in physics and beyond. These efforts are geared toward several envisioned applications, including information processing\cite{Nielsen_Chuang,Leibried_RMP03,Hanson_RMP07,Ladd_Nature10,Buluta_RPP11,Devoret_Science13}, secure communications\cite{Gisin_RMP02,Gisin_NatPhoton07}, and sensing\cite{Bollinger_PRA96,Maletinsky_NN12,Degen_arxiv16}, and enormous progress has been made in engineering and optimizing coherent quantum systems for these purposes. However, decoherence caused by the environment or other factors remains a primary impediment to realizing quantum technologies\cite{Chirolli_AdvinPhys08,Bergli_NJP09,Stanwix_PRB10,Martinis_arxiv14}; overcoming this challenge requires improvements not only in system engineering\cite{Koch_PRA07,Veldhorst_NN14,Muhonen_NatNano14}, but also in how such systems are controlled.

It has been known since the early days of nuclear magnetic resonance that it is possible to design driving fields that suppress adverse effects caused by fluctuations in the system Hamiltonian or in the driving field itself. The simplest example is the Hahn spin echo\cite{Hahn_PR50}, in which a fast ($\delta$-function) $\pi$-pulse is applied halfway through the evolution of a precessing spin, guaranteeing that the spin returns to its initial state at the end of the evolution regardless of the precession rate. This has long been a standard technique to combat inhomogeneous dephasing---the loss of coherence due to variations in precession frequency in spin ensembles. Spin echo and related multi-pulse sequences\cite{Carr_Purcell,Meiboom_Gill,Haeberlen,Wimperis_JMR94,Viola_PRA98,Khodjasteh_PRL05,Uhrig_PRL07,Merrill_Wiley14} have also been widely employed to mitigate other types of decoherence such as environmental noise fluctuations\cite{Bluhm_NP11,Kawakami_PNAS16,Malinowski_NatNano17}. Much work has been done to extend dynamical decoupling to not only preserve the state of the system, but to also cancel errors while performing operations on the system (dynamical gate correction)\cite{Goelman_JMR89,Khodjasteh_PRL10,vanderSar_Nature12,Jones_NJP12,Low_PRA14,Low_PRX16,Kabytayev_PRA14,Soare_NP14, Wang_NatComm12,Kestner_PRL13,Wang_PRA14,Barnes_SciRep15,Calderon_PRL17}.

Although these dynamical decoupling methods have been broadly successful, there are many systems, especially in the context of quantum information technologies, where they exhibit substantial drawbacks. This is because the highly idealized pulse waveforms needed, i.e., $\delta$-functions or square pulses, can be challenging to generate in systems where the dynamics occurs on nanosecond timescales, pushing the limits of current waveform generators, for which minimal achievable rise times are on the scale of 100 ps to 1 ns. In addition, the pulse amplitude is bounded from above by either physical constraints or by the need to avoid overheating the system. These factors, combined with the need to perform operations with an unprecedented level of accuracy, makes standard dynamical decoupling methods inadequate in many systems. For example, quantum computing typically requires operation infidelities below the $10^{-3}$ level, and distortions due to finite rise time and amplitude restrictions become problematic\cite{Merrill_Wiley14}. In addition, building control sequences from a very restricted set of pulse shapes leaves few tunable parameters and leads to unnecessarily long sequences that may compete with other decoherence or loss mechanisms that become important on longer timescales. Smooth pulses can be generated numerically using optimal control techniques such as GRAPE\cite{Khaneja_JMR05,Fouquieres_JMR11} or using continuous analogs of composite sequences such as CORPSE\cite{Merrill_Wiley14,Jones_NJP12}, however such methods yield locally optimal pulses that often exhibit complicated shapes, and obtaining globally optimal quantum controls necessitates the use of analytical methods.

In this paper, we develop a systematic method to obtain all possible driving fields that implement dynamical decoupling and dynamically corrected gates for a resonantly driven qubit. We do this by developing a geometrical framework in which the qubit evolution is mapped onto a curve lying within a two-dimensional plane, where the curvature of the curve determines the shape of the driving pulse, and the length of the curve equals the pulse duration. We show that any closed curve corresponds to an evolution in which the leading-order error vanishes, while any closed curve with zero net area yields a driving field that cancels up to second-order errors. Higher-order error cancellations also admit geometrical interpretations within this construction. We give explicit examples of smooth pulses that perform dynamical decoupling or dynamically corrected gates to arbitrarily high order, and we show that these pulses significantly outperform square and $\delta$-function dynamical decoupling pulses when experimental pulse limitations are taken into account. We also demonstrate the effectiveness of our solutions in the presence of time-dependent $1/f$ noise. In addition, we have written a program, {\sc DDdraw}, that allows the user to create new dynamically correcting pulses simply by drawing curves by hand with a mouse.

\section*{Results}

\subsection{Theoretical setup.}

We focus on the case of a driven qubit described by the following Hamiltonian:
\beq
\mathcal{H}(t)=\frac{\Omega(t)}{2}\sigma_z +\delta \beta \sigma_x.\label{ham}
\eeq
Here, $\Omega(t)$ is the control field, while $\delta\beta$ is an unknown stochastic fluctuation. $\mathcal{H}(t)$ is one of the most widely used models of decoherence due to slow noise\cite{Cywinski_PRB08,Martins_PRL16,Barnes_PRB16}. We can view it as describing a qubit with a tunable energy splitting given by $\Omega(t)$, with $\delta\beta$ interpreted as a fluctuating transverse field. Alternatively, we can think of $\Omega(t)$ as the amplitude of a resonant, monochromatic driving field, in which case $\delta\beta$ represents a fluctuation in the detuning caused by fluctuations in the qubit energy splitting which give rise to inhomogeneous dephasing. In the second case, $\mathcal{H}(t)$ is the Hamiltonian in a frame which rotates with the driving field (in the rotating wave approximation). This Hamiltonian could also describe two driven levels within a larger Hilbert space. For example, the dynamics of three-level lambda systems can often be mapped to an effective driven two-level system.\cite{Press_Nature08,Greilich_NatPhys09,Economou_PRL07}

We parametrize the evolution operator according to
\begin{equation}U(t)=\left(
\begin{array}{cc}
u_{1}(t) & -u_{2}^*(t) \\
u_{2}(t) & u_{1}^*(t) \\
\end{array}
\right).
\end{equation}
It is generally not possible to solve the Schr$\ddot{\hbox{o}}$dinger equation, $i \dot{U}(t)=\mathcal{H}(t)U(t)$, to obtain $u_1$ and $u_2$ analytically aside from a few special choices of $\Omega(t)$. These special choices include the cases of $\delta$-function pulses and square pulses, which is the primary reason why most dynamical decoupling sequences are based on these ideal waveforms. There exist methods for solving this equation more generally using a type of reverse-engineering\cite{Barnes_PRL12,Barnes_PRA13,Barnes_SciRep15}, but these techniques do not apply for the form of $\mathcal{H}(t)$ we are considering here.  Although we cannot obtain closed-form solutions in general, we can obtain solutions as a power series in $\delta\beta$\cite{Wang_NatComm12}:
\begin{equation}
\begin{split}
u_{1}(t) &= e^{-i \phi(t)/2}(g_0(t) - g_2(t)\delta \beta^2 + g_4(t) \delta \beta^4 - ...),\\
u_{2}(t) &= -ie^{i \phi(t)/2}(g_1^*(t) \delta \beta - g_3^*(t) \delta \beta^3 + ...),\label{expansion}\\
\end{split}
\end{equation}
where $\phi(t){=}\int_0^t \Omega(\tau)d \tau$ is the qubit rotation angle, and the coefficients of $\delta \beta$ obey a recurrence relation:
\begin{equation}
g_n(t)=\int_0^t e^{i \phi(\tau)}g_{n-1}^*(\tau)d \tau,\label{recursion}
\end{equation}
with $g_0(t)=1$. To obtain robust quantum operations with operation time $T$, we must find functions $\phi(t)$ (from which we obtain the driving field via $\Omega(t)=\dot\phi$) which yield vanishing error coefficients, $g_m(T)=0$, for all $m\le n$, where $n$ is the desired degree of robustness. Since $g_m$ is complex, this gives two real constraints on $\Omega(t)$ for each odd $m$. For even $m$, there is only one real constraint because unitarity requires $|u_1|^2+|u_2|^2=1$, which from Eq.~\eqref{expansion} implies that at any time $t$,
\beq
\Re[g_n]{=}{\sum_{k=1}^{n/2-1}}(-1)^{k-1}\Re[g_k g_{n-k}^*]-\frac{(-1)^{n/2}}{2}|g_{n/2}|^2.\label{realeven}
\eeq
Thus for even $n$, $\Re[g_n(T)]=0$ is automatically satisfied if $g_m(T)=0$ for all $m\le n/2$, and we are left with one real independent constraint at this order: $\Im[g_n(T)]=0$. For a sequence of $\delta$-function $\pi$ pulses, $e^{i\phi(\tau)}=\pm1$ is the associated characteristic function\cite{Cywinski_PRB08}, and it is easy to show that $g_n(T)=0$ for spin echo, CPMG, and other well known sequences. In what follows, we will show that there exists a geometrical framework that allows us to find all possible driving fields that satisfy these constraints order by order.

\subsection{Geometrical framework and first-order error cancellation.}

\begin{figure}
\centering
\includegraphics[width=0.4\columnwidth]{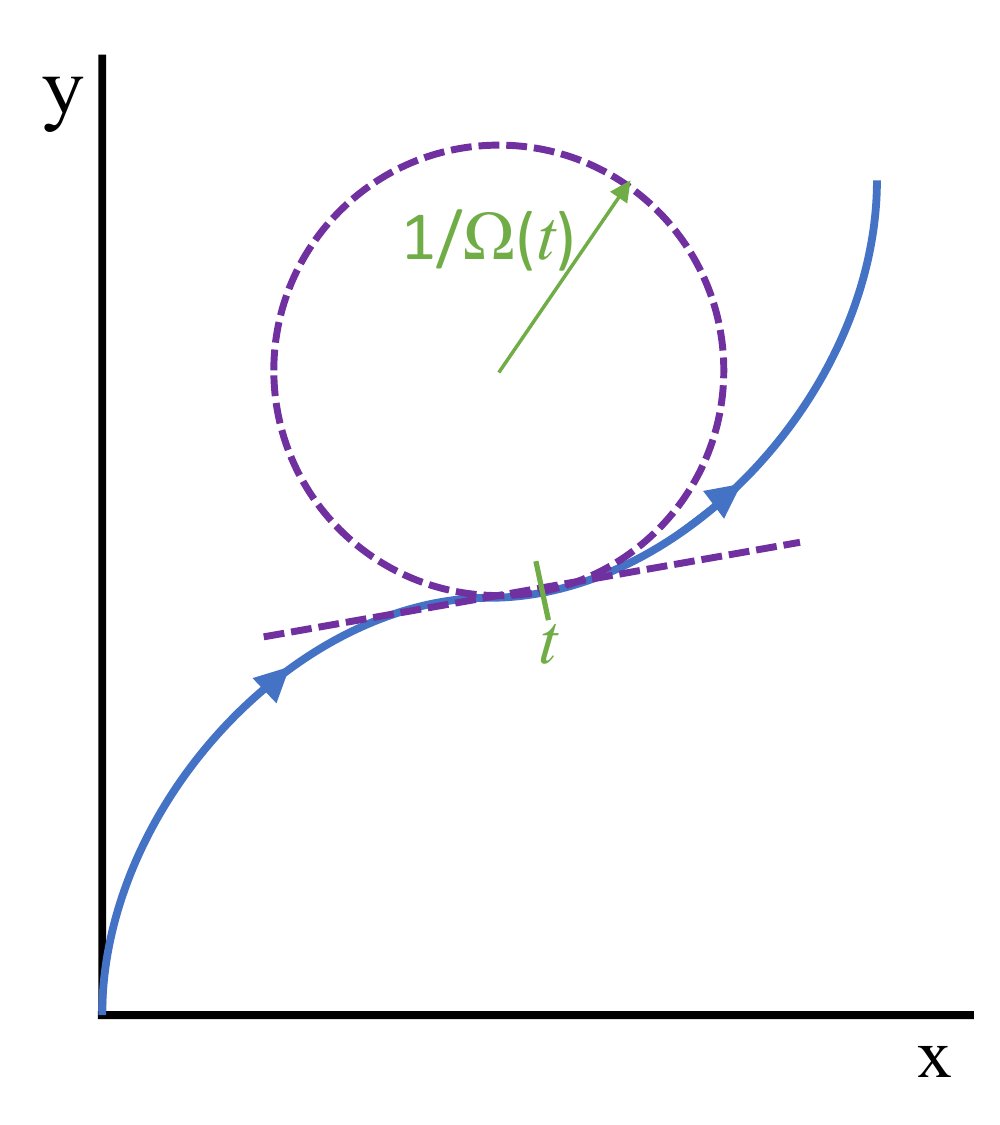}
\caption{The leading-order error coefficient, $g_1(t)$, as a curve in the complex plane. Distance along the curve is parameterized by evolution time $t$, while the driving field $\Omega(t)$ is the curvature at time $t$.\label{fig:schematic}}
\end{figure}

We begin by considering the first-order constraint, $g_1(T)=0$. The solution space of this constraint can be constructed by first parametrizing this function in terms of plane Cartesian coordinates:
\beq
g_1(t) = x(t)+i y(t).\label{g1eqn}
\eeq
Since the time derivative of $g_1(t)$ is a pure phase [Eq.~\eqref{recursion}], we find the following restriction between $x(t)$ and $y(t)$:
\begin{equation}
\dot{x}^2+\dot{y}^2=1.
\label{eq:res}
\end{equation}
The error coefficient, $g_1(t)$, can thus be viewed as a curve in the plane spanned by $x$ and $y$ which starts at the origin at time $t=0$: $x(0)=y(0)=0$, and with slope $\dot x^2(0)+\dot y^2(0)=1$ (see Fig.~\ref{fig:schematic}). There is a one-to-one correspondence between these plane curves and possible driving fields which follows from plugging Eq.~\eqref{g1eqn} into the left-hand-side of Eq.~\eqref{recursion} and differentiating both sides twice:
\beq
\Omega(t)=\dot x\ddot y-\dot y\ddot x=\frac{\dot x\ddot y-\dot y\ddot x}{(\dot x^2+\dot y^2)^{3/2}}.
\eeq
The above formula for $\Omega(t)$ has a surprisingly simple geometric interpretation: it is precisely equal to the signed curvature, $\kappa$, of the plane curve, which is defined to be the proportionality coefficient between the derivative of the curve's tangent vector and its normal vector: $\dot {\bm V}(t)=\kappa(t) \bm{N}(t)$. Since the curvature is defined independently of parametrization, we may express it in terms of a general parameter $\lambda$ along the curve:
\beq
\kappa(\lambda)=\frac{x'y''-y'x''}{(x'^2+y'^2)^{3/2}},\label{curvature}
\eeq
where the prime denotes differentiation with respect to $\lambda$. We are thus free to use any parametrization we choose to define the plane curve. Once we have made this choice, we can extract the driving field: $\Omega(t)=\kappa(\lambda(t))$, where the mapping between $\lambda$ and $t$ is determined from Eq.~\eqref{eq:res}:
\beq
t=\int_0^\lambda d\mu \sqrt{[x'(\mu)]^2+[y'(\mu)]^2}.
\eeq
Notice that the integral on the right hand side measures distance along the curve, meaning that time $t$ can be thought of as the arc-length parametrization of the curve. The driving field is thus the curvature of the plane curve expressed as a function of arc length (see Fig.~\ref{fig:schematic}). This also means that the evolution time $T$ is equal to the total length of the curve. Also notice that the curvature, and hence the driving field, remain invariant under rigid rotations and translations of the curve. Thus, although $\dot g_1(0)=e^{i\phi(0)}=1$ would seem to imply that the initial part of the curve must be tangential to the $x$ axis, it is actually not necessary to enforce this property. 

\begin{figure}
\centering
\includegraphics[width=0.6\columnwidth]{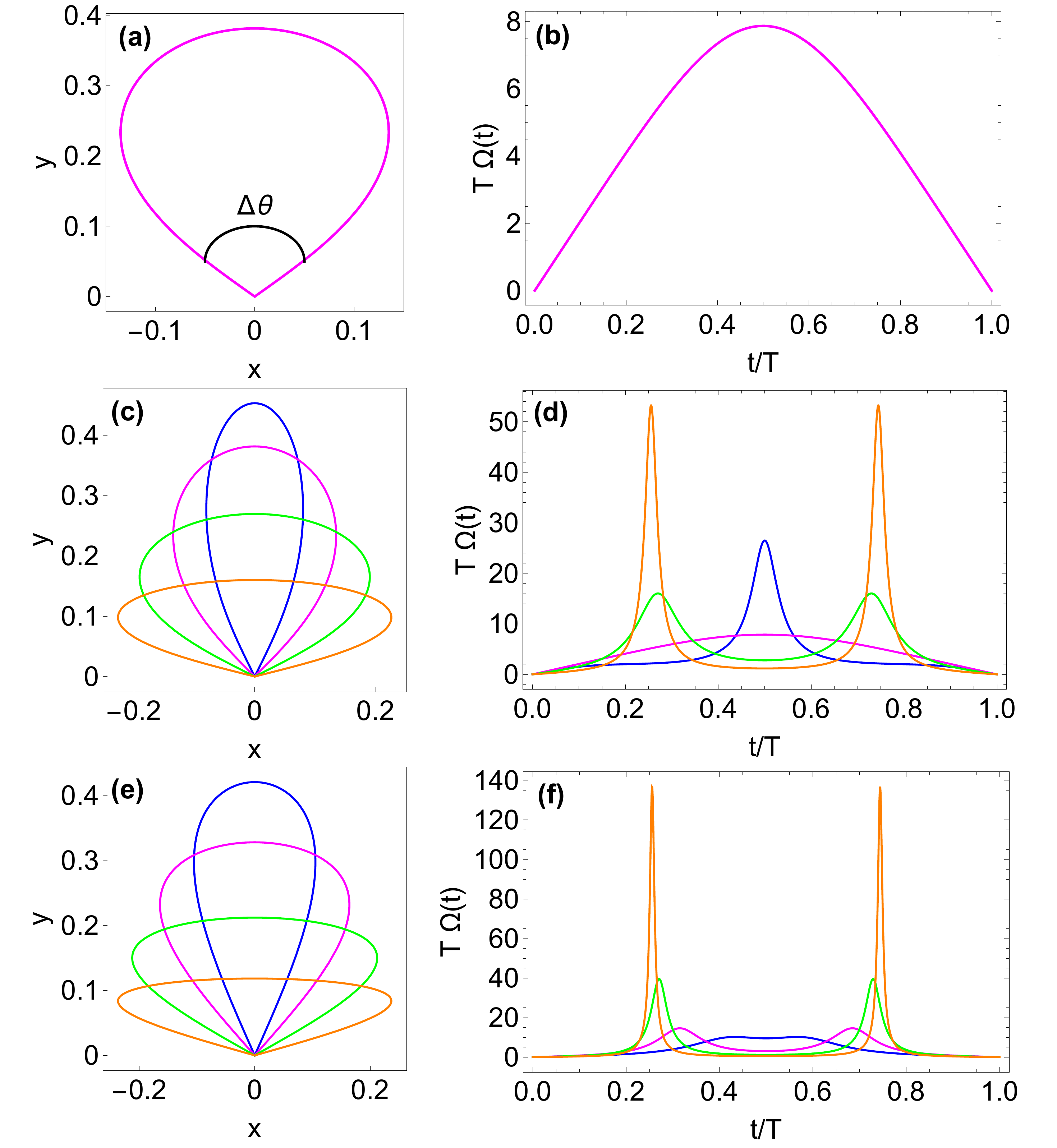}
\caption{Pulses which cancel the first-order error. (a) Example curve with $\Delta\theta=\pi/2$ and (b) corresponding pulse that implements a $3\pi/2$ rotation about the $z$ axis. (c) Modified Bernoulli half-lemniscates, Eq.~\eqref{bernoulli}, and (e) modified Gerono half-lemniscates, Eq.~\eqref{gerono},  with opening angles $\Delta\theta=\pi-2\hbox{arccot}(\alpha)$ for $\alpha=0.5,1,2,4$. All curves have been rescaled to have unit length. (d,f) Corresponding pulses that implement $z$ rotations with angle $\phi(T)=\Delta\theta+\pi$.\label{fig:firstorder}}
\end{figure}

\begin{figure}
\centering
\includegraphics[width=\columnwidth]{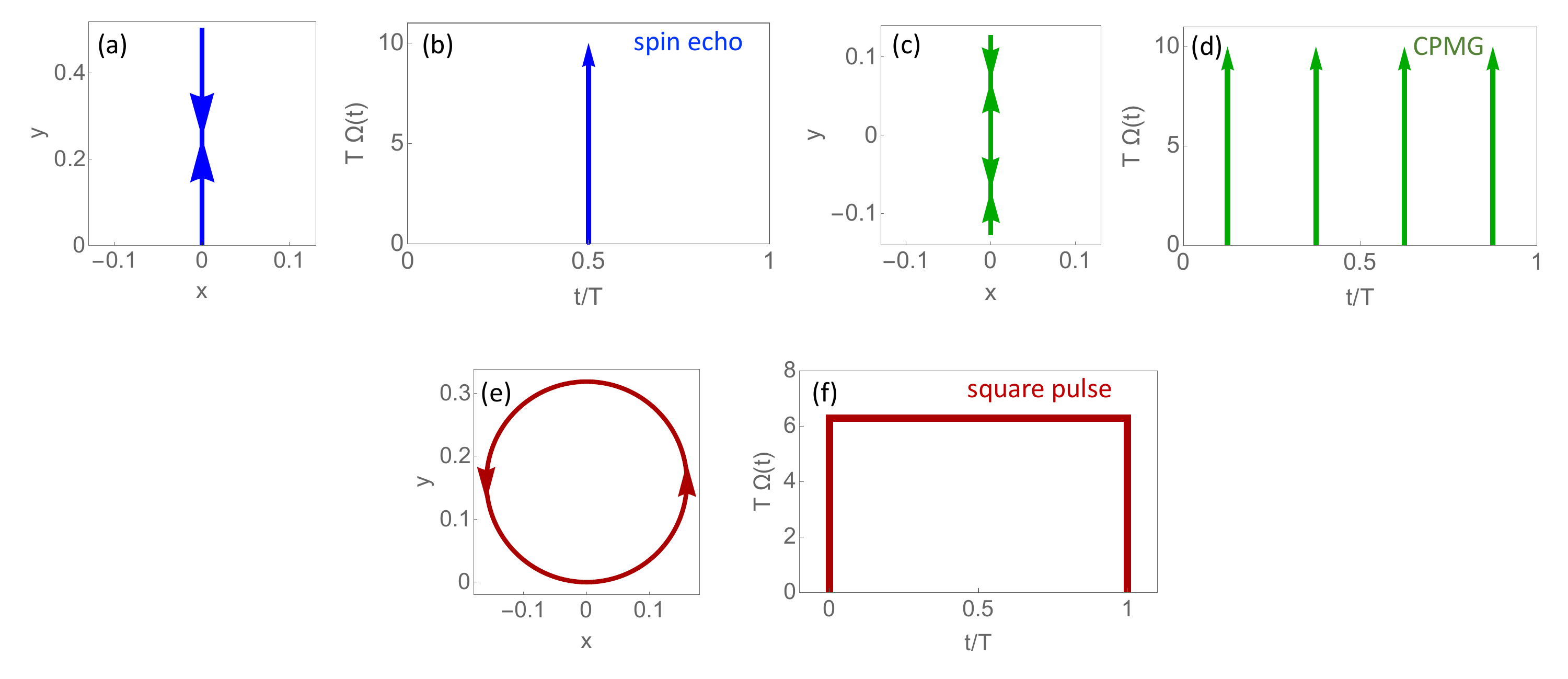}
\vspace{-0.5cm}
\caption{$\delta$-function and square pulses in the geometric framework. A straight line that retraces itself once (a) corresponds to a $\delta$-function $\pi$-pulse (b) that implements Hahn spin echo. A straight line that retraces itself multiple times, starting and ending at the origin (c), corresponds to a CPMG sequence (d). A full circle (e) maps to a single square pulse that implements first-order dynamical decoupling.\label{fig:squaredelta}}
\end{figure}

Requiring the first-order error to vanish at $t{=}T$ is equivalent to imposing the simple condition that the curve is closed: $x(\Lambda)=y(\Lambda)=0$, where $\Lambda\equiv\lambda(T)$. Also, the net rotation angle, $\phi(T)$, is determined by the angle, $\Delta\theta$, subtended by the curve at the origin:
\beq
\phi(T) = \int_0^Tdt\Omega(t)=\arctan\left(\frac{y'}{x'}\right)\bigg|_0^\Lambda=\Delta\theta+\pi.
\eeq
This is illustrated in Fig.~\ref{fig:firstorder}(a) for an example curve with $\Delta\theta=\pi/2$. The corresponding pulse, which implements a $z$-rotation by angle $\phi(T)=3\pi/2$, is shown in  Fig.~\ref{fig:firstorder}(b). The above formula for $\phi(T)$ is valid modulo $2\pi$ because of the multi-valued nature of the inverse tangent. Here, it is important to emphasize that {\it any} closed curve that starts and ends at the origin and which subtends the same angle, $\Delta\theta$, at the origin yields a pulse that implements the same qubit rotation while canceling the leading-order error. Conversely, any pulse which accomplishes these tasks corresponds to a plane curve of this form. Note that in this framework, a single spin echo pulse is represented by the limiting case in which the closed curve collapses on itself to form a straight line segment extending away from the origin a finite distance $r$, as shown in Fig.~\ref{fig:squaredelta}(a),(b); the $\delta$-function $\pi$ pulse corresponds to the sharp turning point at $r$. More general $\delta$-function pulse sequences also correspond to collapsed line segments, with one turning point per pulse (see Fig.~\ref{fig:squaredelta}(c),(d)). Square pulses correspond to circular arcs, while a full circle yields a square pulse that implements first-order dynamical decoupling (see Fig.~\ref{fig:squaredelta}(e),(f)). This geometrical interpretation not only makes it easy to visualize the complete solution space of the first-order constraint, it also makes it easy to systematically determine optimal pulses for a given rotation. This is because the driving field is proportional to the curvature, providing a simple way to visualize and to impose constraints on the smoothness of the pulse. In addition, the fact that the evolution time is equal to the length of the curve allows for optimization of the operation time given smoothness constraints. 

This mapping between robust pulses and closed planar curves is reminiscent of a similar picture in which dynamical decoupling can be interpreted in terms of closed vector paths in the Lie algebra associated with the Hamiltonian\cite{Merrill_Wiley14}. Unlike the present work, however, there is no known method to find the smooth pulses associated with these closed paths, and one is typically restricted to considering piecewise-constant pulses that perform robust identity operations. Interestingly, closed-loop conditions are also known to arise for other types of Hamiltonians, for example ones describing coupled qubit-oscillator dynamics\cite{Hayes_PRL12,Green_PRL15}. 

We can construct simple, explicit examples of smooth pulses which cancel the first-order error by modifying the well known lemniscates (figure-eight curves) of Bernoulli:
\beq
x(\lambda)=\frac{\alpha\sin(2\lambda)}{3+\cos(2\lambda)},\qquad y(\lambda)=\frac{2\sin\lambda}{3+\cos(2\lambda)},\label{bernoulli}
\eeq
and of Gerono:
\beq
x(\lambda)=(\alpha/2)\sin(2\lambda),\qquad y(\lambda)=\sin\lambda.\label{gerono}
\eeq
In both cases, we choose the maximal value of $\lambda$ to be $\Lambda=\pi$ in order to retain only half the lemniscate, giving a closed curve that subtends an angle $\Delta\theta<\pi$ at the origin. We have included the free parameter $\alpha$ in order to make this angle, and hence the rotation angle $\phi(T)$, adjustable: $\Delta\theta=\pi-2\hbox{arccot}(\alpha)$. The example shown in Fig.~\ref{fig:firstorder}(a,b) corresponds to a modified Bernoulli curve with $\alpha=1$. Fig.~\ref{fig:firstorder}(c-f) shows additional examples of half-lemniscates and the corresponding pulses (obtained from the curvature, Eq.~\eqref{curvature}) taken from both the Bernoulli and Gerono families of curves. Together, these solutions give, for each choice of rotation angle $\phi(T)$, two different pulses which implement the same rotation while canceling the leading-order error. Of course, all possible smooth pulses which achieve the same task can be generated from curves qualitatively similar to those shown in Fig.~\ref{fig:firstorder}(c,e). Notice that if we keep the pulse time fixed, then implementing larger rotation angles requires sharper peaks in the pulse, a fact which is clear from the geometrical construction since the plane curve is forced to have sharper bends as $\Delta\theta$ increases.

\begin{figure}
\centering
\includegraphics[width=0.6\columnwidth]{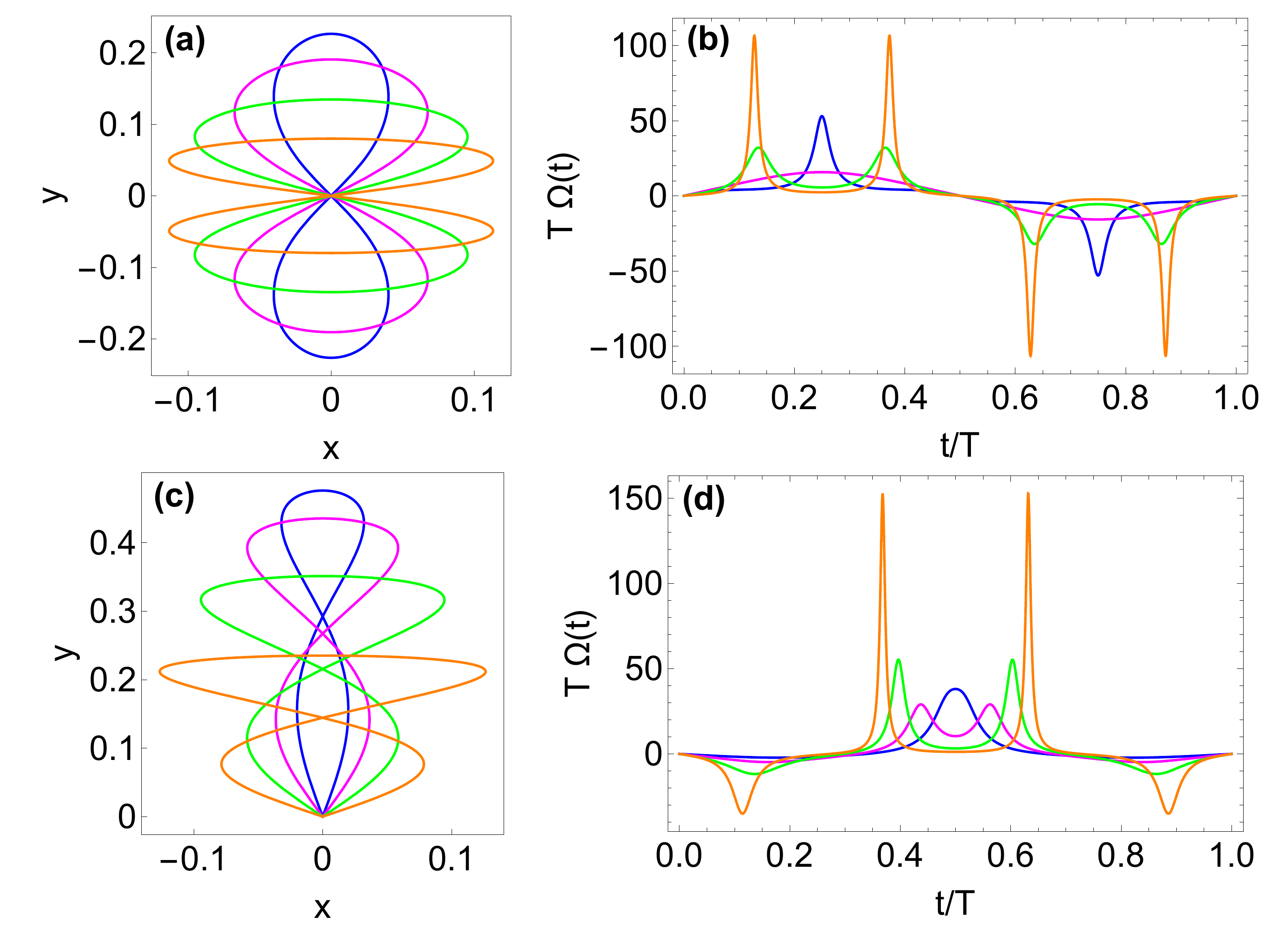}
\caption{Pulses which cancel error to second order. (a) Modified Bernoulli lemniscates, Eq.~\eqref{bernoulli}, for $\alpha=0.5,1,2,4$. All curves have been rescaled to have unit length. (b) Four different pulses corresponding to plane curves in (a) that implement dynamical decoupling ($\phi(T)=0$). (c) Zero-area curves from Eq.~\eqref{2ndordercurves} with $a=1/4,1/2,1,2$ and $b=-6a$. (d) Pulses corresponding to curves in (c) that perform different rotations with angle determined by Eq.~\eqref{2ndorderangle}.\label{fig:secondorder1}}
\end{figure}

\subsection{Second-order error cancellation and {\sc DDdraw}.}

Because of the recursive nature of the error constraints evident in Eq.~\eqref{recursion}, requiring higher-order errors to cancel can be interpreted as imposing additional constraints on the shape of the plane curve. Consider the second-order constraint, $g_2(T)=0$. From Eq.~\eqref{realeven} we know that $\Re[g_2(T)]=0$ if the first-order error vanishes, $g_1(T)=0$, i.e., if the plane curve is closed. The remaining second-order error can be written as
\beq
\Im[g_2(T)]=\int_0^\Lambda d \lambda\left[x'(\lambda)y(\lambda)-y'(\lambda)x(\lambda)\right].
\eeq
Remarkably, this integral is exactly twice the area enclosed by the plane curve, meaning that the second-order error vanishes if and only if the net area enclosed by the curve is zero. Note that the sign of the area is determined by the direction of the winding of the curve, so that the clockwise and counter-clockwise parts of the curve enclose areas of opposite sign.  Thus we find that pulses which cancel both first- and second-order errors are in one-to-one correspondence with closed plane curves that enclose a vanishing net area. Full leminiscates such as those of Gerono and Bernoulli, Eqs.~\eqref{gerono}, \eqref{bernoulli}, satisfy these requirements, provided we now take $\Lambda=2\pi$. The resulting curves and corresponding pulses are shown in Fig.~\ref{fig:secondorder1}(a),(b). Notice that $\Omega(t)<0$ in the second half of the evolution as follows from the negative curvature of one of the lemniscate lobes. This behavior can be interpreted as a sudden $\pi$ phase shift  in the driving field. An important difference compared to the first-order case is that now, each of these pulses implements dynamical decoupling, i.e., an identity operation with $\phi(T)=0$. This is because the angle subtended by the initial and final legs of the plane curves is equal to $\pi$ regardless of the value of $\alpha$. We see from the figure that it is easy to obtain arbitrarily many smooth pulses that perform dynamical decoupling up to second order.

\begin{figure}
    \centering
    \includegraphics[width=0.6\columnwidth]{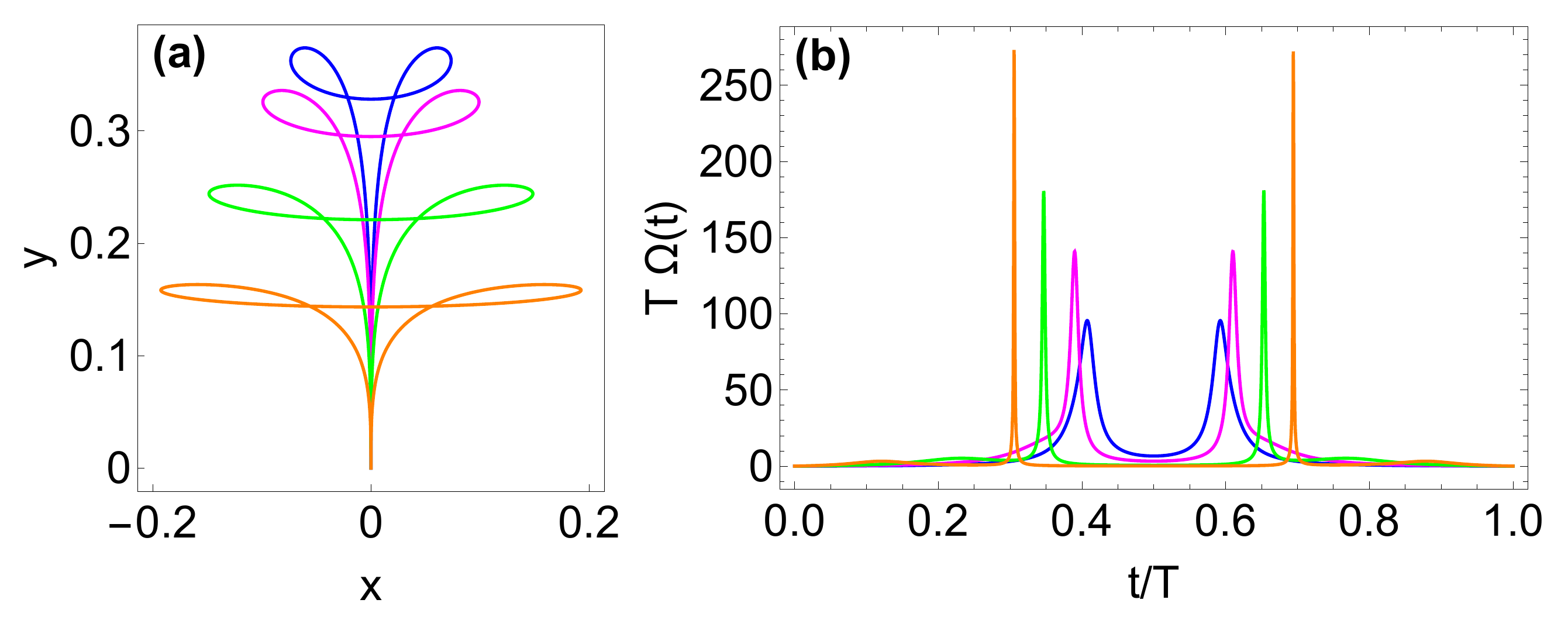}
    \includegraphics[width=0.35\columnwidth]{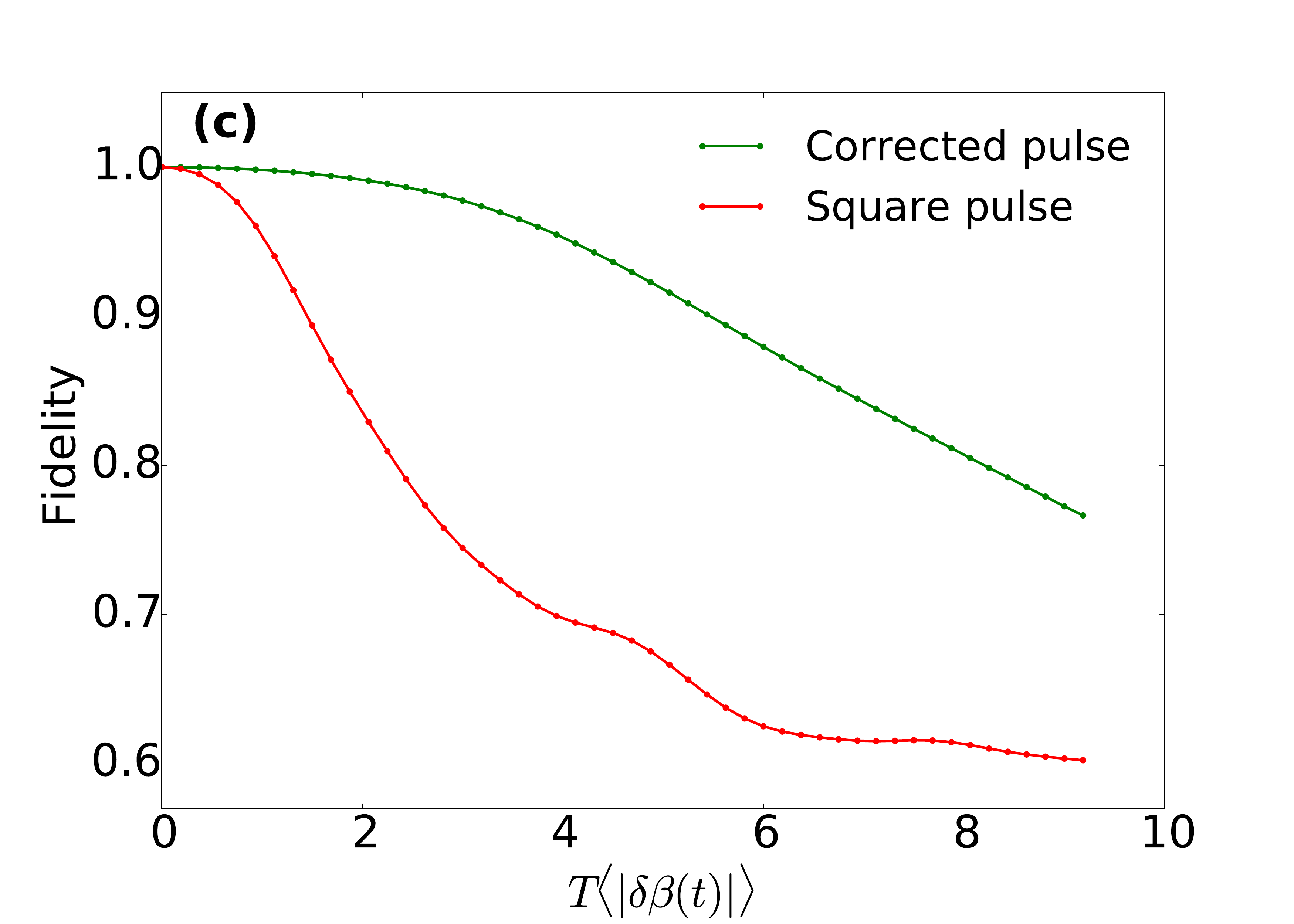}
    \caption{(a) Plane curves from Eq.~\eqref{2ndordercurves} with $a=2/3,1,2,4$ and $b=-a$. (b) Corresponding pulses, all of which implement $3\pi$ rotations while canceling up to second-order errors. (c) Fidelity versus time-averaged noise strength of $1/f$ noise for $3\pi$ pulse with $a=2/3$ compared with square pulse that implements the same rotation with the same duration $T$.\label{fig:secondorder2}}
    \end{figure} 

In order to perform nontrivial rotations while canceling errors up to second order, we need to consider more general zero-area curves that exhibit a kink at the origin. We introduce one family of such curves, parametrized as 
\begin{equation}
	\begin{split}
	x(\lambda)&=[a+b\cos(\lambda)]\sin(\lambda),\\
	y(\lambda)&=\lambda(2 \pi-\lambda)+(4+b/a)[\cos(\lambda)-1],\label{2ndordercurves}
	\end{split}
\end{equation}
with $\Lambda=2\pi$, and where the rotation angle is given by 
\begin{equation}
	\phi(T) = 2\arctan\left(\frac{a+b}{2\pi}\right)+\pi.\label{2ndorderangle}
\end{equation}
Restricting attention to pulses that start and end at zero, $\Omega(0)=\Omega(T)=0$, requires the parameters $a$ and $b$ to be related through either $b=-a$ or $b=-6a$. The former condition produces $3\pi$ rotations while the latter allows for a wide range of rotation angles. Examples of curves with $b=-6a$ and their corresponding pulses are shown in Fig.~\ref{fig:secondorder1}(c),(d). Fig.~\ref{fig:secondorder2}(a) shows examples of curves with $b=-a$, with the associated pulses given in Fig.~\ref{fig:secondorder2}(b). Since our cancellation constraints, Eq.~\eqref{recursion}, are derived assuming quasistatic noise in which $\delta\beta$ is an unknown but constant variation in the Hamiltonian, it is important to consider the performance of the resulting pulses in the presence of time-dependent noise fluctuations. Perhaps the most ubiquitous type of classical time-dependent noise in the context of quantum technologies is $1/f$ noise\cite{Chirolli_AdvinPhys08,Medford_PRL12,Dial_PRL13,Martinis_arxiv14}. Fig.~\ref{fig:secondorder2}(c) shows the operation fidelity\cite{Bowdrey_PLA02} for the pulse corresponding to $a=2/3=-b$ (blue curve in Fig.~\ref{fig:secondorder2}(b)) as a function of noise strength. For comparison, the figure also shows the performance of a square pulse with the same duration. It is clear that even for time-dependent noise, pulses derived from Eq.~\eqref{recursion} outperform naive pulses over a wide range of noise strengths if the noise is sufficiently slow. Further details regarding the numerical simulation can be found in the Methods section.

\begin{figure}
\centering
\includegraphics[width=0.6\columnwidth]{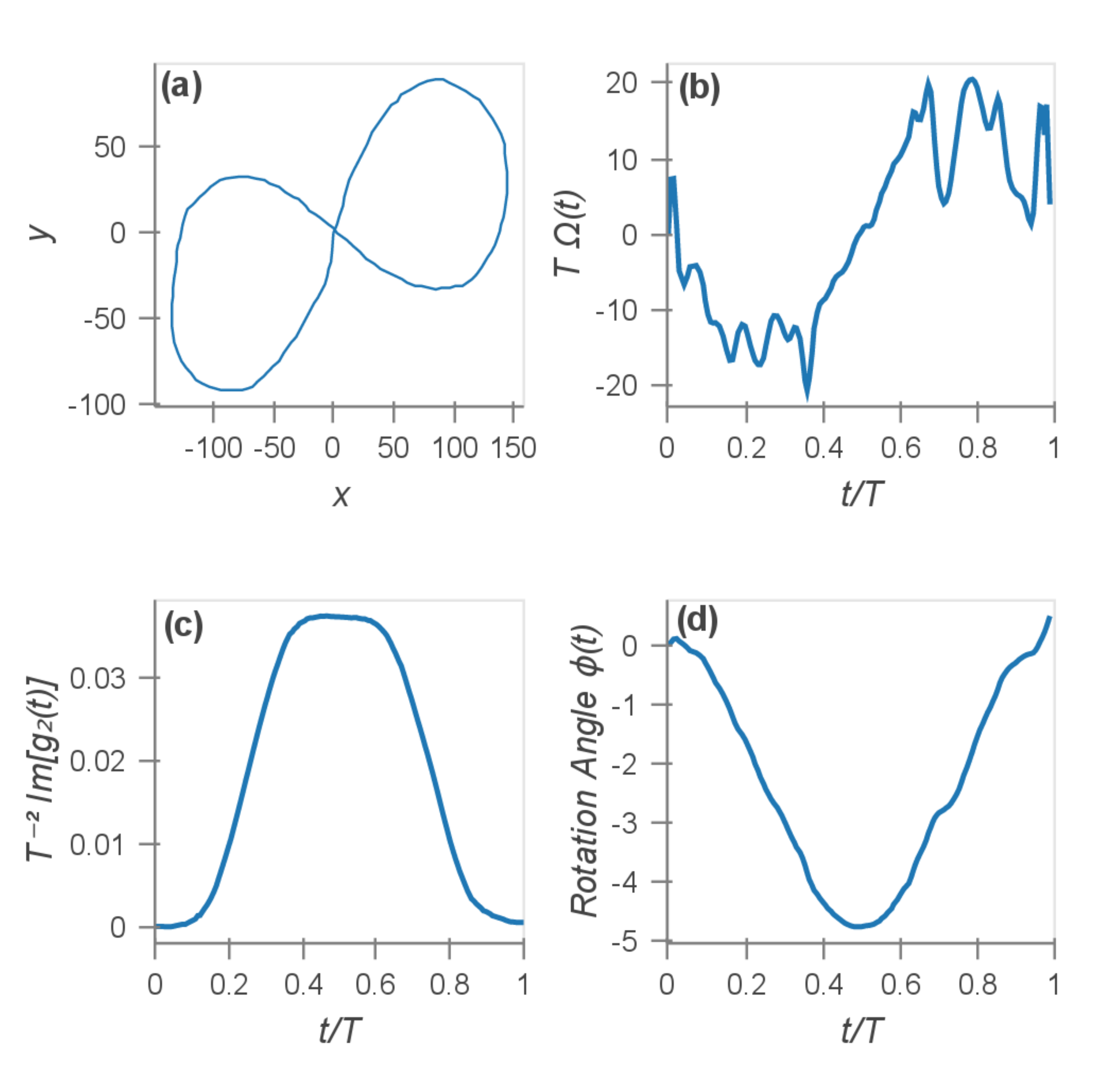}
\caption{Demonstration of plane curve-drawing program {\sc DDdraw}. (a) Hand-drawn plane curve generated using our {\sc DDdraw} program and (b) corresponding pulse. The second-order constraint (c) and rotation angle (d) for this example show that the hand-drawn curve produces a dynamical decoupling pulse that cancels errors up to second order.\label{fig:DDdraw}}
\end{figure}

To further illustrate the simplicity of our geometrical approach to dynamical decoupling and dynamically corrected gates, we have written a program called {\sc DDdraw} \cite{dddraw} that allows the user to draw plane curves by hand using a computer mouse. {\sc DDdraw} displays the corresponding pulse, curve area (second-order cancellation constraint), and rotation angle in real time, making it easier to tailor the pulse as desired and to ensure that the errors are cancelled up to second order. Fig.~\ref{fig:DDdraw} shows an example of a second-order dynamical decoupling pulse generated from a hand-drawn plane curve (Fig.~\ref{fig:DDdraw}(a)) using {\sc DDdraw}.

\begin{figure}
    \centering
    \includegraphics[width=0.4\columnwidth]{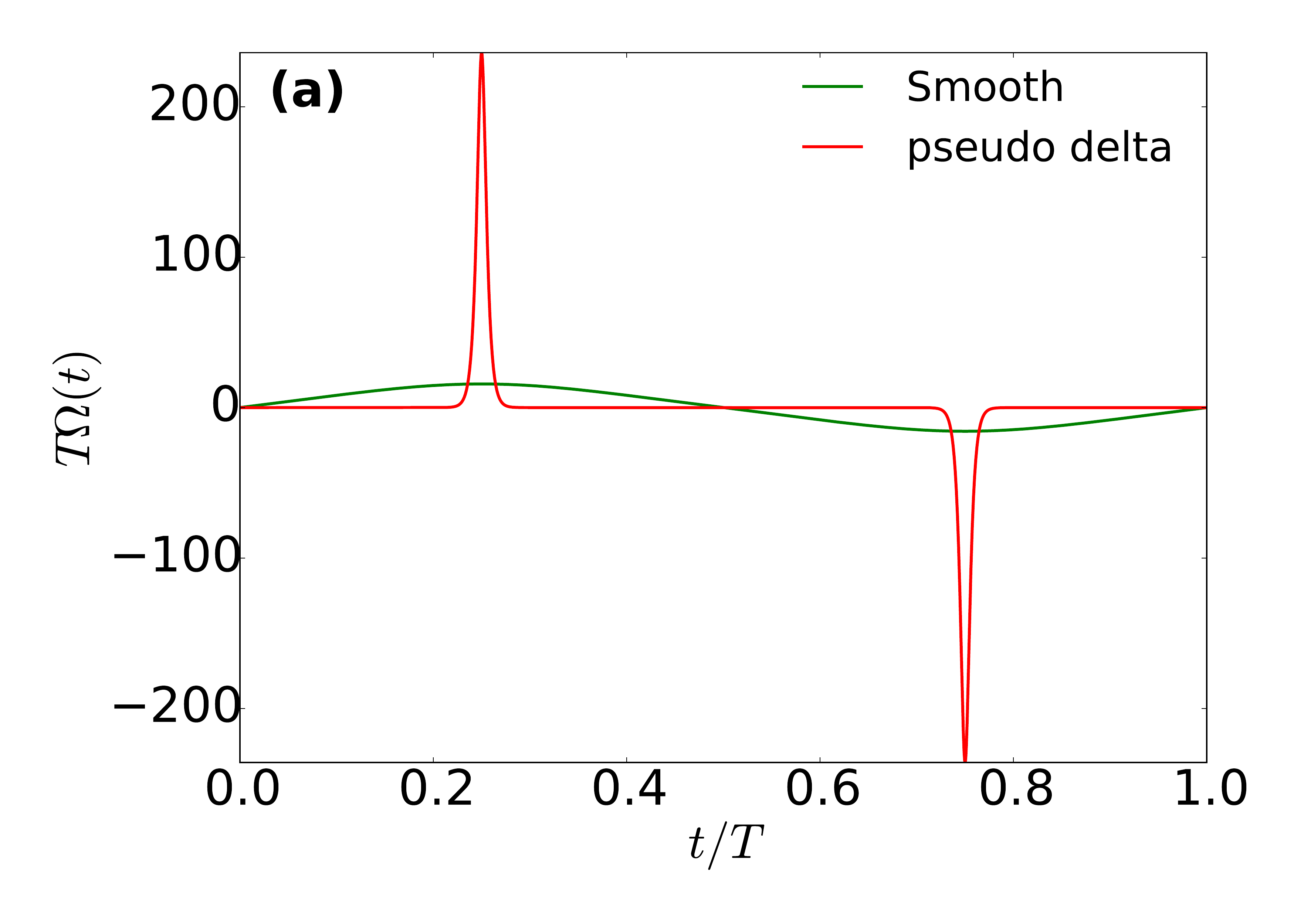}
    \includegraphics[width=0.4\columnwidth]{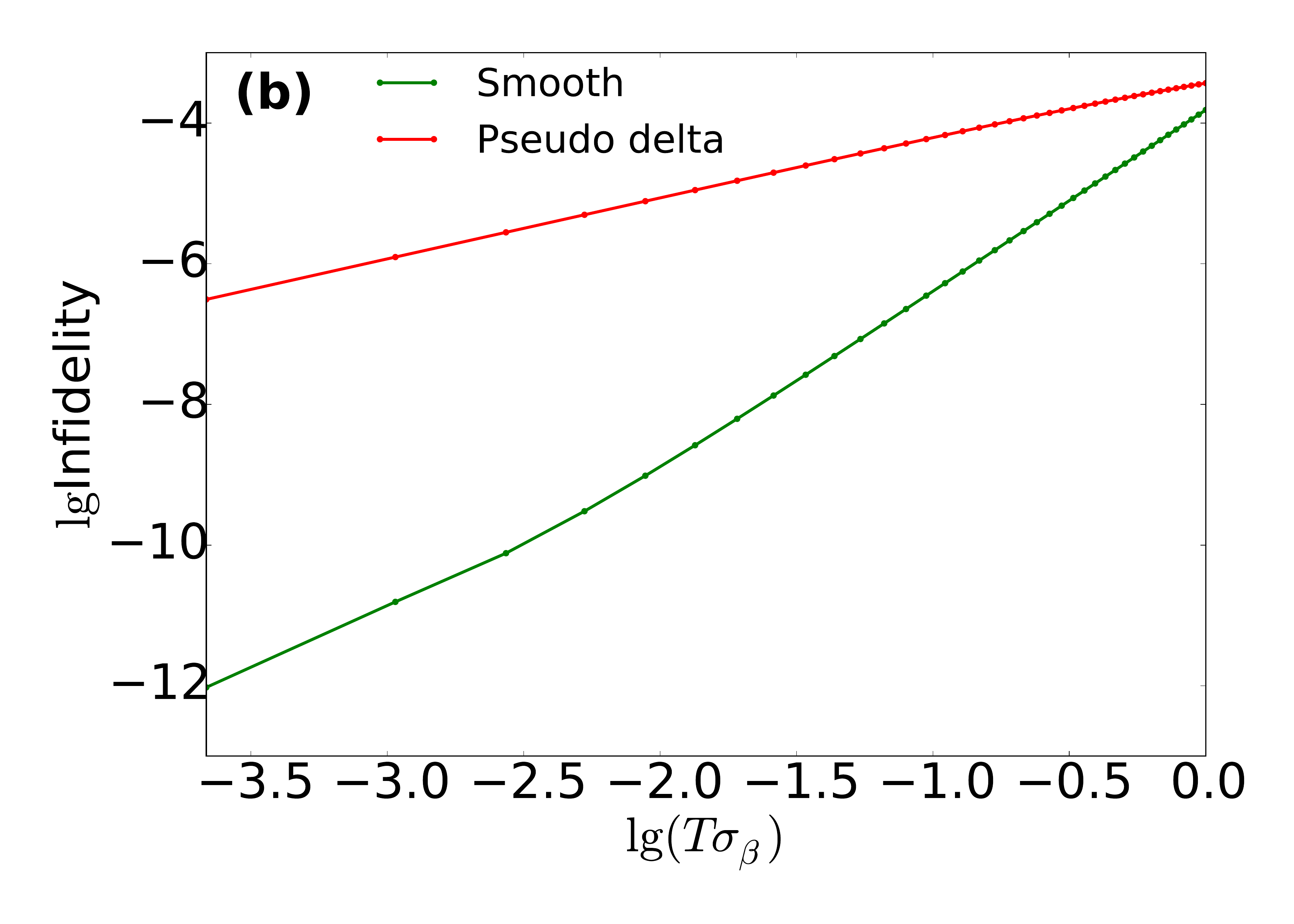}
    \caption{(a) Pseudo $\delta$-function 2-pulse CPMG (red) with finite rise time and amplitude restrictions, and smooth pulse (green) in units of total evolution time $T$. Both control fields are designed to perform identity operations while cancelling transverse noise to second order. (b) Log$_{10}$ of the infidelity versus transverse noise strength for the control fields shown in (a).\label{fig:CPMGcomparison}}
\end{figure} 

\begin{figure}
    \centering
    \includegraphics[width=0.4\columnwidth]{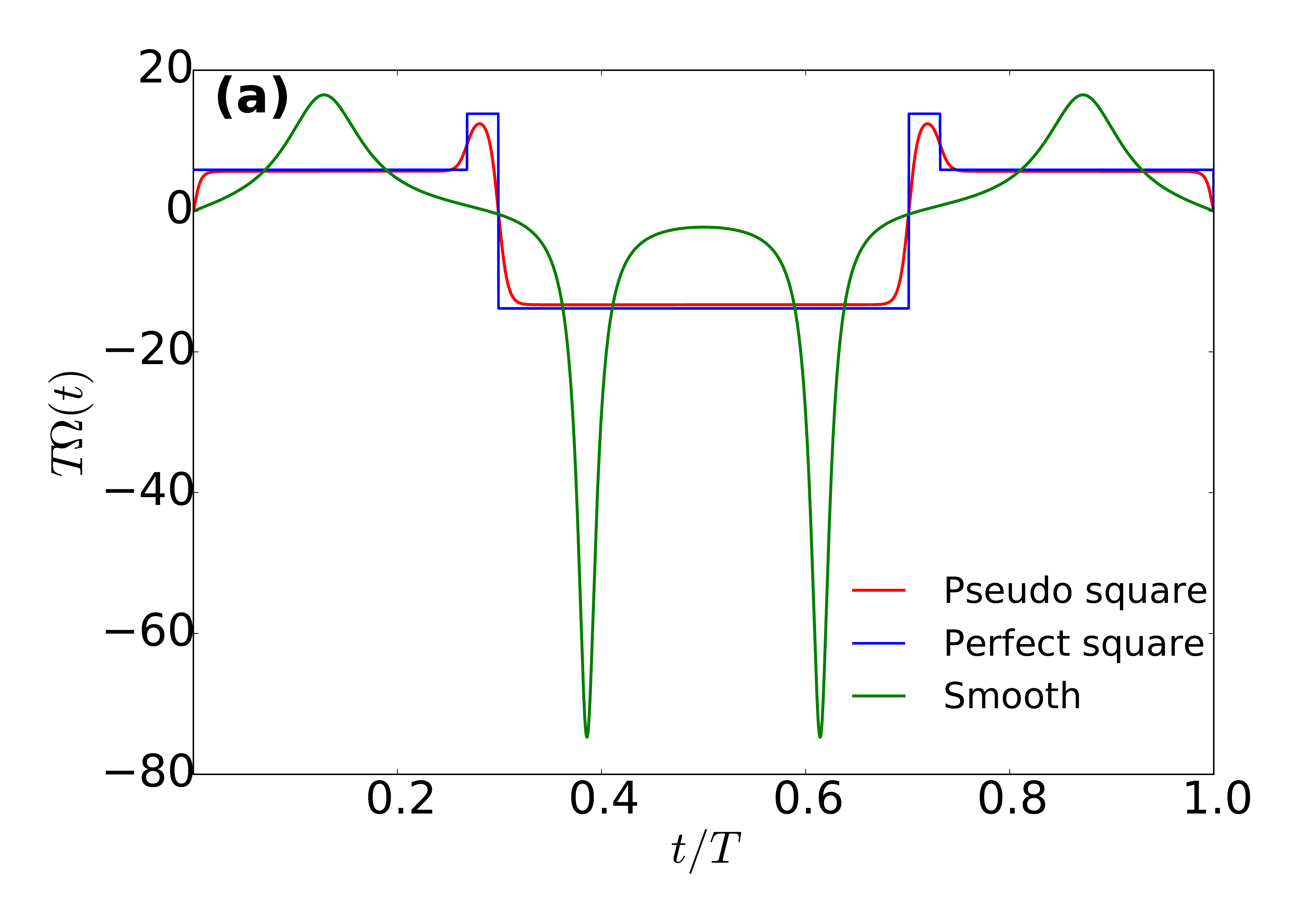}
    \includegraphics[width=0.4\columnwidth]{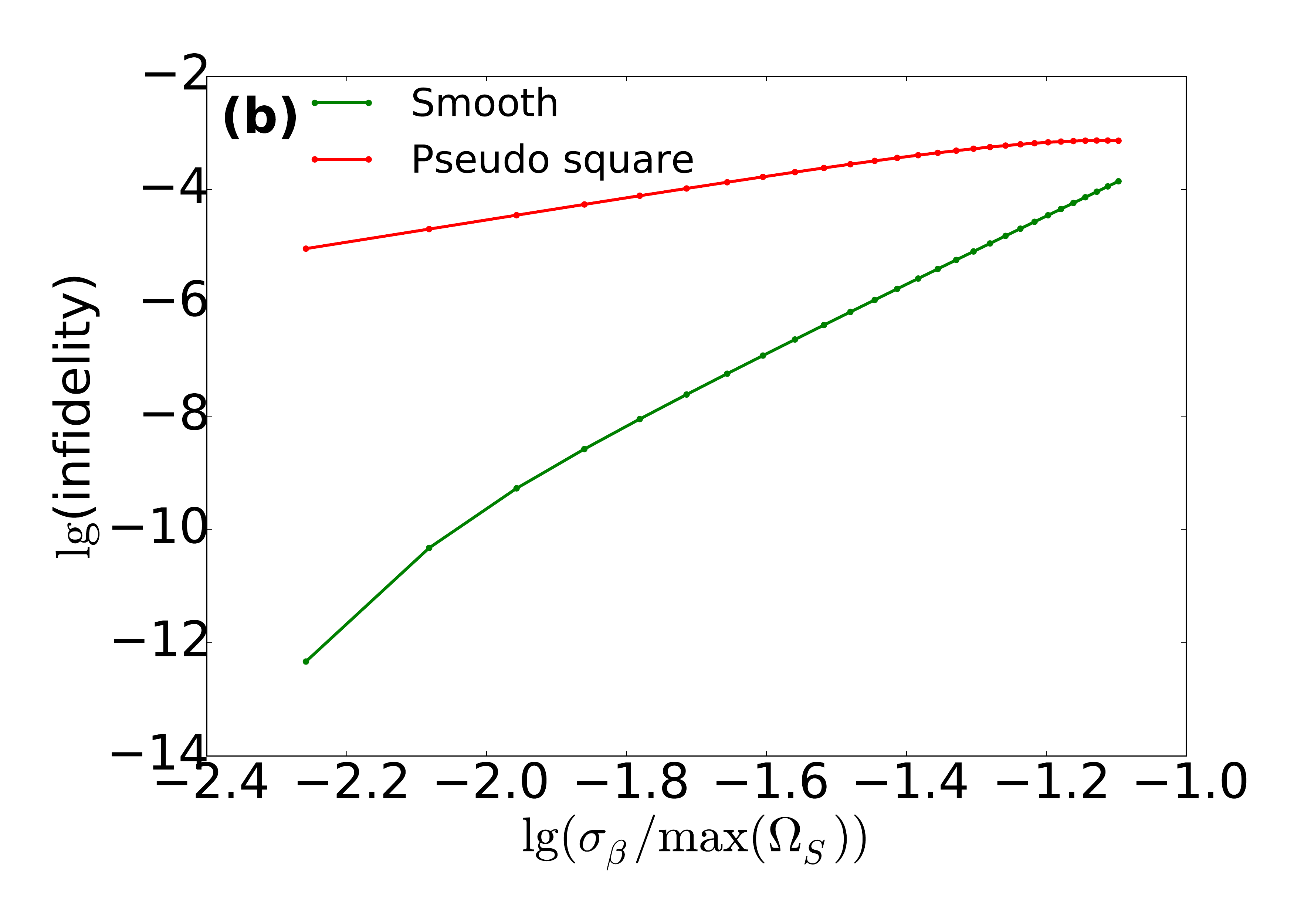}
    \caption{(a) Composite square pulse (blue) and its pseudo-square approximation (red) with finite rise time restrictions imposed, and smooth pulse (green) in units of total evolution time $T$. Both control fields are designed to perform $\pi/2$ rotations about $z$ while cancelling transverse noise to second order. (b) Log$_{10}$ of the infidelity versus transverse noise strength for the control fields shown in (a). The noise strength is measured in units set by the maximal amplitude of the ideal composite pulse, $\Omega_S$.\label{fig:squarecompare}}
\end{figure}

\subsection{Comparison to imperfect square and $\delta$-function pulses.}

The advantage of smooth pulses over square or $\delta$-function pulses comes from experimental limitations in pulse generation. Microwave pulses produced from arbitrary waveform generators cannot vary arbitrarily quickly; typical rise time limits are on the scale of 100 ps to 1 ns. This leads to unavoidable errors when square or $\delta$-function pulses are attempted in experiments, and these errors can be appreciable for pulse times on the order of nanoseconds. Limitations on pulse amplitude can come from physical constraints on the system Hamiltonian (e.g., voltage-controlled exchange pulses are constrained by the upper bound on the strength of the exchange interaction set by the confinement energy\cite{Petta_Science05}) or from the need to avoid overheating the system (as is the case for e.g., magnetically driven electron spins in quantum dots\cite{Hanson_RMP07}). This leads to further errors in the case of $\delta$-function dynamical decoupling since the pulses may be leveled off at relatively low amplitudes because of these restrictions.

To demonstrate the significance of these effects and the advantage afforded by smooth pulses, we compare our smooth pulses with square and $\delta$-function decoupling pulses subject to both amplitude and rise time restrictions. We first consider the case of pure dynamical decoupling where the goal is to perform an identity operation while canceling noise errors to second order. Ideally, this can be achieved with many different $\delta$-function sequences, but for concreteness we focus on 2-pulse CPMG (which cancels errors to arbitrary order in the case of quasistatic noise). We account for amplitude and rise time restrictions by considering the following ``pseudo'' $\delta$-function sequence made from hyperbolic secant pulses:
\beq
\Omega_{CPMG}(t)=\frac{\gamma}{T}[\hbox{sech}(\gamma(t-T/4))-\hbox{sech}(\gamma(t-3T/4))],
\eeq
where we choose $\gamma=47.2$ so that the area of each pulse is $\pi$, and $T$ is the total evolution time. This 2-pulse sequence is shown in Fig.~\ref{fig:CPMGcomparison}(a). Here, we have chosen the rise time to be 2\% of the total evolution time. In this figure, we also show one of our smooth pulses, the Bernoulli lemniscate with $\alpha=1$ shown in Fig.~\ref{fig:secondorder1}(b), which is also designed to perform an identity operation while canceling second-order errors. In Fig.~\ref{fig:CPMGcomparison}(b) we compare the infidelities for these two types of dynamical decoupling as a function of transverse noise strength. Here, the noise is taken to be quasistatic Gaussian noise, and the strength is characterized by the width $\sigma_\beta$ of the Gaussian distribution. We find that the infidelity is reduced by several orders of magnitude over a broad range of noise strengths when the smooth pulse is used.

Next, we perform a similar comparison, but now for a non-trivial rotation. Ideally, such a rotation can be performed while cancelling noise errors using a composite square pulse such as the one shown in Fig.~\ref{fig:squarecompare}(a). This composite pulse is designed to cancel errors to second order while implementing a $\phi=-\pi/2$ rotation about $z$. We impose rise time restrictions by deforming this composite pulse into a smoothened, pseudo-square version (red curve in the figure). We again take the rise time to be 2\% of the total evolution time for concreteness. The precise form of this composite pulse is
\beq
\Omega_{comp}(t)=f(t) \sum_{i=1}^5 a_i \left[1-\tanh \left(b \left(-l_i+t-t_{0i}\right)\right)\right] \left[\tanh \left(b \left(l_i+t-t_{0i}\right)\right)+1\right],
\eeq
where
\beq
f(t)=\frac{\tanh(4bt)[\tanh(4b(t-T))-1]+\tanh(4b(t-T))+\tanh(4bT)}{\tanh(4bT)-\tanh(2bT)[2+\tanh(2bT)]},
\eeq
and the parameters $t_{0i}=(x_i+x_{i-1})/2$ and $2l_i=x_i-x_{i-1}$ give respectively the center and the duration of the $i$’th square pulse, which starts and ends at $t=x_{i-1}$ and $t=x_i$, respectively. Here, we choose the parameters $x/T=(0,0.268,0.299,0.701,0.732,1)$, $aT=(1.422,3.357,-3.357,3.357,$ $1.422)$, and $bT=110.581$, which yields the pulse shown in Fig.~\ref{fig:squarecompare}(a). This figure also shows one of our smooth error-suppressing pulses which implements the same rotation and cancels noise to second order while obeying the same rise time restriction. This pulse is obtained from the family of curves introduced in Eq.~\eqref{2ndordercurves} by choosing $a=2\pi/5$ and $b=-6a$. The advantage of this pulse over the composite one is evident in Fig.~\ref{fig:squarecompare}(b), which shows that the infidelity for the smooth pulse is orders of magnitude smaller than that of the composite pulse for a wide range of noise strengths. In quantum computing and related fields, where infidelities typically need to be suppressed to the $10^{-3}$ level or below, this advantage can become very important.

\subsection{3D Minkowski interpretation.}

\begin{figure}
    \centering
    \includegraphics[width=0.4\columnwidth]{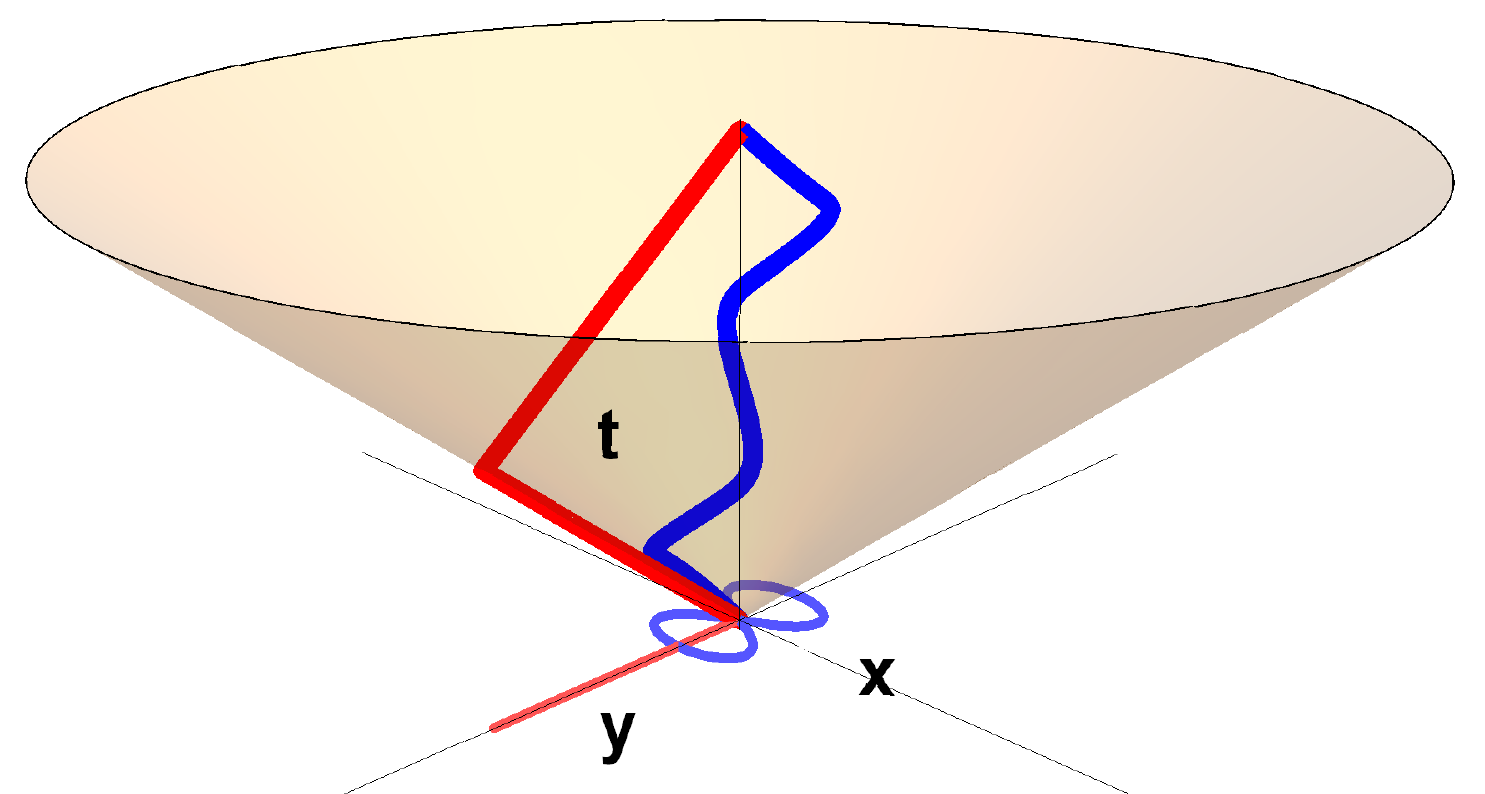}
    \caption{There is a one-to-one correspondence between driving fields and lightlike (null) curves in 3D Minkoswki space. Driving fields which cancel the first-order error correspond to lightlike curves that start and end at the origin. The 2D plane curves are the projections of these lightlike curves onto the $xy$ plane. The blue line is the lightlike curve for the Bernoulli lemniscate, Eq.~\eqref{bernoulli}, with $\alpha=2$. The red line is the lightlike curve corresponding to a spin echo pulse (a $\delta$-function $\pi$-pulse applied halfway through the evolution). The 2D projections of both lightlike curves are also shown. The lightlike curves must lie on or within the lightcone with apex at $x=y=t=0$.\label{fig:3dMinkowski}}
    \end{figure} 

\begin{figure}
    \centering
    \includegraphics[width=0.8\columnwidth]{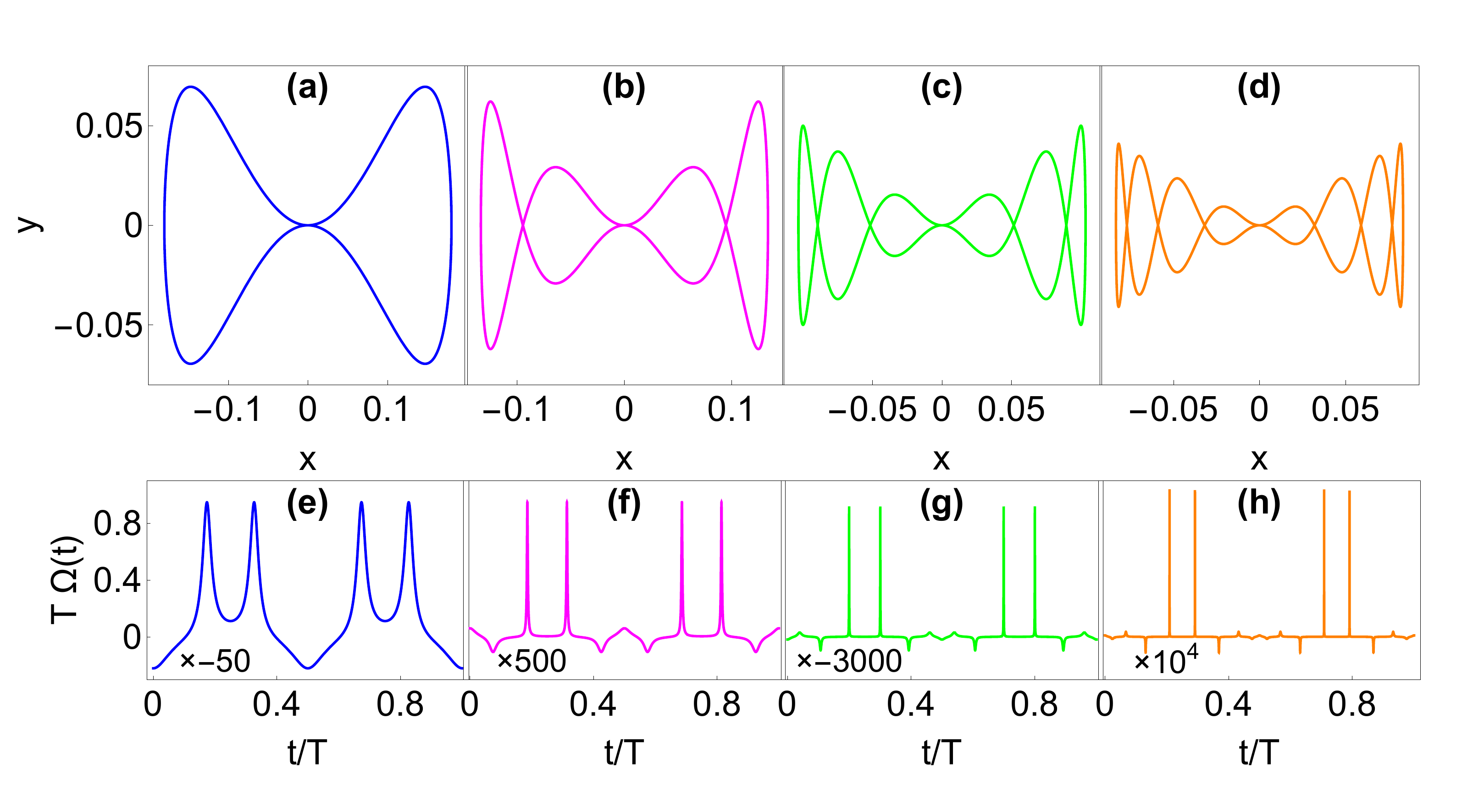}
    \caption{Higher-order dynamical decoupling. (a-d) Plane curves from Eq.~\eqref{deformedgerono} with $m=2,4,6,8$. All curves have been rescaled to have unit length. (e-h) Corresponding pulses which implement identity operations while canceling up to first-, third-, fifth-, and seventh-order errors, respectively.\label{fig:higherorder}}
    \end{figure} 

The relationship between the evolution time and the distance along the curve has an additional geometrical interpretation: the plane curve can be viewed as a lightlike (or null) curve in 3D Minkowski space. This fact can be seen by rewriting Eq.~\eqref{eq:res} as $-dt^2+dx^2+dy^2=0$, which is the infinitesimal proper distance along a lightlike curve. In this interpretation, any $\Omega(t)$ which cancels the first-order error maps onto a lightlike curve which starts and ends at the origin of 3D Minkowski space. Such a curve is necessarily confined within a lightcone with apex at $x=y=t=0$, as shown in Fig.~\ref{fig:3dMinkowski}. This lightcone is a reflection of the quantum speed limit, which refers to the minimum time it takes to evolve from one quantum state to another\cite{Mandelstam_JPhys45,Bhattacharyya_JPA83,Margolus_PD98,Giovannetti_PRA03,Barnes_PRA13,Arenz_arxiv17}. For a two-level system, the quantum speed limit is saturated for zero driving, $\Omega(t)=0$, which corresponds to a straight line in the plane, or equivalently a lightlike curve lying on the 3D lightcone. This is consistent with our earlier observation that $\delta$-function pulse sequences correspond to line segments in the plane: in the 3D Minkowski picture, the evolution between $\delta$-pulses is represented by lightlike curves (straight lines parallel to the lightcone), with a sharp bend for each pulse. This is illustrated in Fig.~\ref{fig:3dMinkowski} for the case of spin echo.

\subsection{Cancelling higher-order errors.}
The higher-order error-cancellation constraints give further restrictions on the shape of the plane curves. In the Methods section,
 we show that the constraints at arbitrary order $n$ are built from two basic classes of integrals:
\beq
p_k(\lambda)=\int_0^\lambda d\mu r^k\theta',\quad q_k(\lambda)=\int_0^\lambda d\mu r^k\theta'e^{i\theta},\label{pandq}
\eeq
where $r$ and $\theta$ are plane polar coordinates: $re^{i\theta}=x+iy$, and $k$ is an integer. For example, third-order cancellation introduces the new constraint, $q_3(\Lambda)=0$, while for fourth-order cancellation, we need $p_4(\Lambda)=0$. These can be understood as requiring that certain signed volumes bounded by the plane curves must vanish. Consider for instance the constraint $p_k(\Lambda)=0$. Since the area enclosed by the plane curve can be written as $\tfrac{1}{2}\int_0^\Lambda d\mu r^2(\mu)\theta'(\mu)$, we recognize that $p_k(\Lambda)$ is proportional to the volume lying between the plane curve area and the function $F(r)=r^{k-2}$ over the plane. The constraints $q_k(\Lambda)=0$ admit a similar interpretation. In the Methods section, 
we show that the following family of deformed Gerono lemniscates parametrized by even integer $m$,
\beq
x(\lambda)=\sin\lambda,\qquad y(\lambda)=\frac{1}{2}\sin(m\lambda)\sin\lambda.\label{deformedgerono}
\eeq
satisfy the error constraints up to order $m-1$. These curves and their associated pulses are shown in Fig.~\ref{fig:higherorder}, where it is evident that higher-order errors are canceled by inserting additional self-intersections in the curve. For example, the curve shown in Fig.~\ref{fig:higherorder}(a) yields a pulse (Fig.~\ref{fig:higherorder}(e)) that cancels only the first-order error error, while the curve in Fig.~\ref{fig:higherorder}(b), which contains two additional self-intersections, gives a pulse (Fig.~\ref{fig:higherorder}(f)) that cancels up through third order. The self-intersections of the curves allow integral constraints of the form $p_{2n}(T)=0$ to be satisfied since the lobes of the curve on either side of the intersection then have areas of opposite sign. Each additional lobe that appears in Fig.~\ref{fig:higherorder}(a-d) has precisely the area need to cancel $p_{2n}(T)$ for all $n<m/2$. Integrals of the type $q_{2n+1}(T)$ for $n<m/2$ cancel due to the inversion symmetry of the curves (see Methods).  
It is apparent from Fig.~\ref{fig:higherorder} that for fixed pulse time, the pulses become sharply peaked as higher-order errors are canceled. This is a consequence of needing to insert more turns and self-intersections in the plane curves in order to cancel higher-order errors, and it suggests that an all-orders error cancellation can only be achieved with $\delta$-function pulses. This is consistent with a no-go theorem\cite{Wang_NatComm12} stating that all-orders error cancellation is not possible with finite pulses implemented within a finite time. 

The above findings also imply that if we keep the maximal pulse amplitude $\Omega_{max}$ fixed (and hence the maximal allowed curvature of the plane curve), then the pulse duration $T$ must increase as the order of the error cancellation is increased. This in turn implies that the pulse bandwidth would decrease since it behaves like $1/T$. For the particular class of error-cancelling pulses shown in Fig.~\ref{fig:higherorder}(e)-(h), we have calculated $\Omega_{max}T$ for the first 6 pulses (corresponding to $m=$2,4,6,8,10,12) in the series (the last of which cancels up to 11th order errors) and found that the maximal bandwidths are $1/T=\Omega_{max}\times[0.02,0.002, 3.6\times10^{-4}, 9.6\times10^{-5}, 3.3\times10^{-5}, 1.4\times10^{-5}]$. This data can be fit to the function $\exp(-a m^b)$ where $a=2.41$ and $b=0.68$, which shows that the maximal bandwidth decreases subexponentially with increasing error-cancellation order.

\section*{Conclusion}

We have developed a geometrical framework that yields all possible driving fields that suppress inhomogeneous dephasing or errors due to qubit energy splitting fluctuations. We have used this framework to construct several explicit examples of smooth driving fields that implement dynamically corrected gates up to second order or dynamical decoupling to arbitrary order, and we have demonstrated that these outperform dynamical decoupling sequences based on $\delta$-function or square pulses when realistic constraints on pulse rise times and amplitudes are taken into account. We have also used this framework to write software called {\sc DDdraw} that allows the user to easily generate new dynamical decoupling pulses by drawing plane curves by hand. The simple way in which the pulse shape is encoded in this geometrical construction (as the curvature of the plane curves) makes it straightforward to find optimal pulses given a set of system-specific constraints, for example on amplitude, slope, or overall pulse duration. Our framework thus provides a general, powerful approach to improving the control of qubit systems.

\section*{Methods}

\subsection{$1/f$ noise simulation.} Here, we provide details regarding our time-dependent $1/f$ noise simulations. We simulate $1/f$ noise as a sum over many random telegraph noise (RTN) sources:
\beq
\delta\beta(t)=\sum_{i=0}^Nw_i\eta_i(t),\label{expanddh}
\eeq
where $\eta_i(t)$ represents a stochastic noise fluctuation for a single RTN source, $N+1$ is the total number of RTN sources, and $w_i$ is a weight factor that we must determine. The above expansion implies that
\bea
&&\langle \delta \beta(t)\delta \beta(0)\rangle=\sum_{i=0}^N\sum_{j=0}^Nw_iw_j\langle\eta_i(t)\eta_j(0)\rangle\nn\\
&&=\sum_{i=0}^N\langle\eta_i(t)\eta_i(0)\rangle w_i^2=\sum_{i=0}^N e^{-2|t|/\tau_i}w_i^2,
\eea
where in the final step, we used that different RTN sources are uncorrelated and that the two-point correlation function of the $i$th RTN source decays exponentially with characteristic timescale defined as $\tau_i$. Taking the Fourier transform of both sides, we find that the noise power spectrum is given by
\beq
S(\omega)=\sum_{i=0}^N\frac{4\tau_i}{4+\omega^2\tau_i^2}w_i^2.
\eeq
We want to determine the RTN parameters such that this spectrum approximates that of $1/f$ noise:
\beq
S(\omega)\approx\frac{A^2}{\omega}.
\eeq
More precisely, this relation should hold in the continuum limit, $N\to\infty$. This will be the case if we choose the $\tau_i$ to be equally spaced:
\beq
\tau_i=\frac{i}{N}(\tau_{max}-\tau_{min})+\tau_{min},\label{defoftaui}
\eeq
and if we choose the weight factor to be
\beq
w_i^2=\frac{A^2}{N\pi}\frac{\tau_{max}-\tau_{min}}{\tau_i}.\label{wivals}
\eeq
Eqs.~\eqref{expanddh}, \eqref{defoftaui}, and \eqref{wivals} are what we use to approximate $1/f$ noise using an ensemble of RTN sources in numerical simulations. For the numerical results shown in the main text, we take $N=10^5$, $\tau_{min}=44T$, $\tau_{max}=441T$, and vary $A$ so that the time-averaged noise strength, $\mean{|\delta\beta(t)|}$, varies from 0 to $9.2/T$, averaging over 50 instances of the noise for each value. The square pulse used in the comparison is taken to have the same duration, $T$, as our smooth error-correcting pulse, as well as the same area of $3\pi$. We use the definition of fidelity given in\cite{Bowdrey_PLA02}.

\subsection{Form of higher-order constraints.}
Starting from the recursion relation, Eq.~\eqref{recursion}, and performing integrations by parts multiple times yields one of four expressions depending on the value of $n\mod4$:
\bea
g_n&=&g_{n-1}^*g_1-g_{n-2}g_2^*+\ldots-g_{(n+2)/2}g_{(n-2)/2}^*\nn\\&&+\int dt g_{n/2}^*{\dot g}_{n/2}, \qquad n=4m+2,\nn\\
g_n&=&g_{n-1}^*g_1-g_{n-2}g_2^*+\ldots+g_{(n+2)/2}^*g_{(n-2)/2}\nn\\&&-\int dt g_{n/2}{\dot g}_{n/2}^*, \qquad n=4m+4,\nn\\
g_n&=&g_{n-1}^*g_1-g_{n-2}g_2^*+\ldots+g_{(n+3)/2}^*g_{(n-1)/2}\nn\\&&-\int dt g_{(n+1)/2}{\dot g}_{(n-1)/2}^*,\qquad n=4m+1,\nn\\
g_n&=&g_{n-1}^*g_1-g_{n-2}g_2^*+\ldots-g_{(n+3)/2}g_{(n-1)/2}^*\nn\\&&+\int dt g_{(n+1)/2}^*{\dot g}_{(n-1)/2},\qquad n=4m+3,
\eea
where $m$ is a non-negative integer. We see that if we assume $g_k(T)=0$ for $k<n$, then the constraint $g_n(T)=0$ reduces to a condition involving two lower-order error coefficients:
\bea
\int_0^T dt g_{n/2}^*{\dot g}_{n/2}&=&0, \qquad \hbox{$n$ even},\nn\\
\int_0^T dt g_{(n+1)/2}^*{\dot g}_{(n-1)/2}&=&0, \qquad \hbox{$n$ odd}.\label{intconstraints}
\eea
Taking $n=2$ and working with plane polar coordinates, $g_1=re^{i\theta}$, we obtain
\beq
g_2(t)=\int_0^td\tau g_1^*{\dot g}_1=\frac{1}{2}r(t)^2+i\int_0^td\tau r^2\dot\theta.
\eeq
We recognize the integral on the right as the area enclosed by the plane curve when $t=T$. Using this result for $g_2$ in Eq.~\eqref{intconstraints} and assuming $r(T)=0$, we find that the third-order constraint simplifies to
\beq
g_3(T)=\frac{4i}{3}\int_0^Tdt r^3e^{i\theta}\dot\theta=0.
\eeq
Similarly, if we also assume $g_2(T)=0$, then cancellation of the fourth-order constraint requires 
\beq
g_4(T)=i\int_0^Tdt r^4\dot\theta=0,
\eeq
while for fifth order we have
\beq
g_5(T)=\frac{8i}{15}\int_0^Tdt r^5e^{i\theta}\dot\theta-\frac{8}{3}\int_0^Tdtr^2\dot\theta\int_0^tdt' r^3e^{i\theta}\dot\theta,\label{5thorder}
\eeq
and for sixth order,
\beq
g_6(T)=\frac{2i}{9}\int_0^Tdt r^6\dot\theta+\frac{16}{9}\int_0^Tdt r^3e^{i\theta}\dot\theta\int_0^tdt' r^3e^{-i\theta}\dot\theta.\label{6thorder}
\eeq
Notice that $g_4(T)$ and $g_6(T)$ are both purely imaginary (the latter assumes $g_3(T)=0$) as expected from Eq.~\eqref{realeven}. The fact that all these constraints consist of integrals of $\dot p_k(t)$, $\dot q_k(t)$ (which are defined in Eq.~\eqref{pandq}) and their complex conjugates arises because the functions $g_n(t)$ are themselves comprised of such terms. This in turn can be proven using induction in conjunction with Eq.~\eqref{recursion}.

\subsection{Dynamical decoupling to arbitrary order.}
In this section, we show that the deformed Gerono lemniscate of order $m$ given in Eq.~\eqref{deformedgerono} solves the error-cancellation constraints up to order $m-1$. First, we show that integrals of the type $p_{2n}(T/2)$ vanish for $m>2n$, where $p_k(t)$ is defined in Eq.~\eqref{pandq}. To do this, we begin by rewriting this integral:
\beq
p_{2n}(T/2) \propto \int_0^\pi  \sin ^{2 n}(\lambda ) \left(\sin ^2(m \lambda )+4\right)^{n-1} \, d\sin(m \lambda)\nn
\eeq
\beq
=\sum_{i=0}^{n-1}a_i\int_0^\pi\sin^{2n}(\lambda)\sin^{2i}(m \lambda) \, d\sin(m \lambda)\nn
\eeq
\beq
=\sum_{i=0}^{n-1}\frac{a_i}{2i+1}\int_0^\pi\sin^{2n}(\lambda)\, d\sin^{2i+1}(m \lambda)\label{rewritep2n}
\eeq
\beq
=\sum_{i=0}^{n-1}\tilde{a}_i\int_0^\pi \sin^{2i+1}(m \lambda)\sin^{2n-1}(\lambda)\cos(\lambda)d \lambda\nn
\eeq
\beq
=\sum_{i=0}^{n-1}\sum_{j=0}^{i}\tilde{a}_i{b_j}\int_0^\pi \sin((2j+1)m \lambda)\sin^{2n-1}(\lambda)\cos(\lambda)d\lambda.\nn
\eeq
In this formula, each term has the form
\begin{equation}
\begin{split}
	I_{2n+1}&=\int_0^{\pi} \sin(2\ell \lambda)\sin^{2n+1}(\lambda)\cos(\lambda)d\lambda\\
	&=-\ell\int_0^\pi \cos(2\ell \lambda)\sin^{2n+2}(\lambda)d \lambda - n I_{2n+1}.
\end{split}
\end{equation}
Thus we get $I_{2n+1}=-\frac{\ell}{n+1}J_{2n+2}$, where we have defined
\begin{equation}
	J_{2n}=\int_0^\pi \cos(2\ell \lambda)\sin^{2n}(\lambda)d \lambda.\label{defJn}
\end{equation}
Using integration by parts, we can also obtain
\begin{equation}
	J_{2n+2} = \frac{-\ell}{n+1}I_{2n+1}+\frac{2n+1}{2n+2}J_{2n},
\end{equation}
so we have in the end
\begin{equation}
	I_{2n+1} = \frac{-(2n+1)\ell}{2(n+1+\ell)(n+1-\ell)}J_{2n}.
\end{equation}
We see from Eq.~\eqref{defJn} that $J_0=0$, so from the relation above it follows that for $n+1<\ell$ we have $I_{2n+1}=0$. If we combine this conclusion with the formula in Eq.~\eqref{rewritep2n}, we find that as long as $m>2n$ the deformed Gerono lemniscate, Eq.~\eqref{deformedgerono}, is able to cancel $p_{2n}(T/2)$. The cancellation of these integrals implies that $p_{2n}(T)$ also vanishes.

The deformed Gerono lemniscates also cancel the integrals $q_{2n+1}(T)$, where $q_k(t)$ is defined in Eq.~\eqref{pandq}. This follows from the curves' inversion symmetry. Every point, $\lambda$, along the curve has an inversion-related partner, $\tilde\lambda$, such that $\theta(\tilde\lambda)=\pi+\theta(\lambda)$, $r(\tilde\lambda)=r(\lambda)$, and $\theta'(\tilde\lambda)=\theta'(\lambda)$. This implies that the contribution to $q_{2n+1}(T)$ from the $x<0$ half of the curve exactly cancels that from the $x>0$ half.

In addition to terms like $p_n(T)$ or $q_n(T)$, the $n$th order error coefficient also contains terms with nested integrals involving $p_k(t)$ and $q_k(t)$ (see e.g., Eqs.~\eqref{5thorder} and \eqref{6thorder}). We do not have a proof that such terms also vanish identically for the deformed Gerono lemniscates, but we have verified this numerically up to seventh order and conjecture that this trend continues to arbitrary order.

\section*{Acknowledgments}
The authors thank Sophia Economou and Ken Brown for helpful comments. This work is supported by the Army Research Office  (W911NF-17-0287) and by the Office of Naval Research (N00014-17-1-2971).

\section*{Author contributions}
JZ and EB conceived the project, performed calculations, and wrote the manuscript. XD assisted with numerical simulations, and AR developed the program {\sc DDdraw}. EB oversaw the project.

\section*{Author Information}
The authors declare that they have no competing financial interests. Correspondence
and requests for materials should be addressed to Edwin Barnes (email: efbarnes@vt.edu).

\end{document}